\documentclass[sigconf,noacm]{acmart}
\AtBeginDocument{%
  }
\setcopyright{none}
\acmISBN{978-1-4503-XXXX-X/2018/06}

\usepackage{makecell}
\usepackage{soul}
\usepackage{multirow}
\usepackage[table]{xcolor}
\usepackage{xeCJK}
\begin{document}

% \title{Exascale Generative Compression of Earth Observation Data on CPU Supercomputers for 500x Storage Reduction}
\title[Transforming the Use of Earth Observation Data: Exascale Training of a Generative Compression Model]{Transforming the Use of Earth Observation Data: \\ Exascale Training of a Generative Compression Model \\ with Historical Priors for up to 10,000× Data Reduction}

\author{
  \begin{tabular}{@{}c@{}}
    {Jinxiao Zhang}$^{2*}$, {Runmin Dong}$^{3*}$\textsuperscript{\textdagger}, {Xiyong Wu}$^{1*}$, {Xihan Huang}$^{1}$, {Shenggan Cheng}$^{4}$, \\
    {Yunkai Yang}$^{3}$, {Zheng Zhou}$^{1}$, {Yunpu Xu}$^{1}$, {Zhaoyang Luo}$^{1}$, {Miao Yang}$^{2}$, {Fan Wei}$^{2}$, {Mengxuan Chen}$^{2}$, \\
    {Yang You}$^{4}$, {Juepeng Zheng}$^{3}$, {Weijia Li}$^{1}$, {Yutong Lu}$^{3,5}$, {Haohuan Fu}$^{1,2,5}$\textsuperscript{\textdagger}
  \end{tabular}
}
\affiliation{
  \begin{tabular}{@{}c@{}}$^{1}$ Institute of Data and Information, Tsinghua Shenzhen International Graduate School \\ 
    $^{2}$Department of Earth System Science, Tsinghua University \quad $^{3}$Sun Yat-Sen University \\ $^{4}$National University of Singapore \quad 
    $^{5}$National Supercomputing Center in Shenzhen \\ 
    $^{*}$Equal contribution, \textsuperscript{\textdagger}Corresponding authors. 
  \end{tabular}
  \country{}
}

% $^{*}$Equal contribution, 

\renewcommand{\shortauthors}{Jinxiao et al.}

\begin{abstract}
   Earth observation is becoming one of the largest data-producing activities in science, yet current pipelines still treat compression as a storage and transmission tool rather than a new way to use data. We present a generative compression framework that learns from historical Earth observation archives and enables on-demand 100× to 10,000× data reduction across downstream tasks. Unlike general visual data, Earth observation repeatedly measures the same evolving planet, making historical-prior learning feasible for extreme compression. To realize this paradigm, we train large generative compression models at exascale on the LineShine Armv9 CPU supercomputer, with co-optimization across model design, kernels, memory hierarchy, runtime, and parallelism. Our implementation sustains 1.54 EFLOP/s and peaks at 2.16 EFLOP/s in end-to-end training. This work shows that historical-prior generative compression can turn Earth observation data into an active, task-adaptive foundation for acquisition, delivery, storage, and scientific use.
  % A clear and well-documented \LaTeX\ document is presented as an
  % article formatted for publication by ACM in a conference proceedings
  % or journal publication. Based on the ``acmart'' document class, this
  % article presents and explains many of the common variations, as well
  % as many of the formatting elements an author may use in the
  % preparation of the documentation of their work.
\end{abstract}

% <concept_id>10010147.10010169</concept_id>
% <concept_desc>Computing methodologies~Parallel computing methodologies</concept_desc>
% <concept_id>10010520.10010521</concept_id>
% <concept_desc>Computer systems organization~Architectures</concept_desc>
% <concept_id>10010147.10010371.10010395</concept_id>
% <concept_desc>Computing methodologies~Image compression</concept_desc>

% \ccsdesc[500]{Computing methodologies~Parallel computing methodologies}
% \ccsdesc[500]{Computer systems organization~Architectures}
% \ccsdesc[500]{Computing methodologies~Image compression}



\maketitle
\setlength{\parskip}{0.4em} 
% https://awards.acm.org/bell/nominations#h-submissions
\section{Justification for ACM Gordon Bell Prize}

We realize the first exascale training system for historical-prior generative compression of Earth observation data, sustaining 1.54 EFLOP/s and peaking at 2.16 EFLOP/s. Beyond compression itself, this work establishes a new foundation for how Earth observation data are encoded, delivered, stored, and used across % downstream tasks.
% \vspace{2cm}

\section{Performance Attributes}

\begin{table}[H]
  \small
  \centering
    \begin{tabular}{|c|c|}
    \hline
    Attributes & Contents \\
    \hline
    Category achievement type & \makecell[c]{\textit{Scalability, Time-to-solution,} \\ \textit{Peak performance}, \textit{Throughput}} \\
     \hline
    Type of method used & \makecell[c]{\textit{Dense Transformer Model}} \\
     \hline
    Results reported & \textit{Whole application including I/O} \\
     \hline
    Precision reported & \textit{BFloat16} \\
     \hline
    System scale & \textit{Measured on full-scale system} \\
     \hline
    Measurement mechanism & \makecell[c]{\textit{Timer, FLOP counts} } \\
    \hline
    \end{tabular}%
    \vspace{-0.2cm}
\end{table}%

\section{Overview of the Problem}

Over the past half century, Earth observation has transformed humanity's long effort to observe and understand the Earth into one of the largest and longest-running scientific projects ever focused on a single evolving system.\footnote{This long-standing motivation resonates with classical ideas across civilizations. In the Chinese tradition, it finds an early expression in the \textit{Dao De Jing (道德经)}: ``Man follows Earth, Earth follows Heaven, Heaven follows the Dao, and the Dao follows nature.''}
Multi-generational satellite programs such as Landsat and Sentinel have created a vast observational archive across time, scale, modality, and geography~\cite{wulder2016global,drusch2012sentinel}. This extraordinary data wealth should enable scientists to study climate change, ecological dynamics, and coastline evolution in new ways. Yet the scale of the archive has made its scientific use increasingly prohibitive: major Earth observation archives now already span hundreds of petabytes~\cite{nasa_earthdata_metrics,sentinel_success_stories_2026}, and for many important problems, the data that must be accessed, moved, stored, and repeatedly analyzed have reached the PB scale. As a result, broad use of this information remains a luxury rather than a norm. The bottleneck in Earth observation is therefore shifting from data acquisition to data usability~\cite{chi2016big}. Addressing this bottleneck would do more than reduce storage or transmission cost; it would create a new, more equitable and platform-like foundation for Earth observation data access, sharing, and scientific innovation~\cite{zhou2021rsims}.

Current Earth observation pipelines remain organized around acquisition, downlink, storage, and distribution, with compression serving mainly as a passive utility for reducing bandwidth and archive cost~\cite{yu2009image,li2024earth,zhou2021rsims}. This design has supported large-scale data collection, but it does not fundamentally change how data are used: scientists must still retrieve, move, store, and repeatedly process massive raw or lightly compressed datasets before analysis. Turning this pipeline into a fundamentally new mode of data use is difficult for three reasons. First, Earth observation archives are enormous, heterogeneous, and continually growing, spanning multiple decades, sensors, spectral bands, resolutions, and operating modes. Second, many scientific and operational tasks impose stringent fidelity requirements, so aggressive reduction must preserve not only visual structure but also physically meaningful information. Third, making historical archives directly usable through a learned model requires training and maintaining that model over global-scale data, which creates an extreme systems challenge in data throughput, model scale, and computation. Existing compression methods therefore alleviate parts of the burden, but they do not yet turn Earth observation archives into an active, task-adaptive interface for data access and scientific use.

Recent advances in large language models suggest that large-scale compression can become a pathway to usable knowledge when the underlying data already provide a structured encoding of the world. Natural language is such a medium: it is compositional, cross-scale, and accumulated through generations of human observation. Raw visual data, by contrast, generally do not admit such direct large-scale compression into a unified model, because pixel space is far less structured and has much higher effective degrees of freedom. Earth observation may represent a rare exception. Unlike arbitrary visual streams, it repeatedly measures the same evolving planet over decades, yielding strong regularities across geography, seasonality, surface structure, and multispectral response~\cite{chi2016big}. As a result, global Earth observation archives are not merely massive collections of images, but a historical record of one persistent evolving system observed from many sensors and scales. This makes it plausible to ask whether historical Earth observation data can be compressed into a generative model that does not simply reconstruct pixels, but serves as a learned, task-adaptive prior for how the Earth is queried and used, rather than only as passive reference information~\cite{du2025earth+}.

Motivated by this perspective, we develop a historical-prior generative compression framework for Earth observation data. Rather than using compression only to reduce storage and transmission cost, our approach trains a large generative model on global historical archives so that long-term geographic and spatiotemporal priors become directly usable for task-adaptive data reduction. This enables a new mode of interaction with Earth observation data, in which users need not rely exclusively on raw PB-scale archives, but can instead access model-mediated representations matched to different downstream objectives. To make this paradigm practical, we combine large-scale archive construction, generative model design, and exascale training on the LineShine Armv9 CPU supercomputer. The resulting system supports on-demand up to 10,000× data reduction while reframing how Earth observation data can be encoded, delivered, stored, and used.

\section{Current State of The Art}

The state of the art can be examined along three connected dimensions. First, existing compression methods for Earth observation define the boundary of what can be recovered from the current observation and its transmitted bitstream. Second, recent remote sensing foundation models show that global historical EO archives can support the learning of strong priors, but not yet in a form tailored to extreme-bitrate recovery. Third, large-scale AI training practice shows what system performance is possible at scale, while leaving open the question of how to make this particular workload practical on an emerging CPU supercomputer. Taken together, these three dimensions reveal a common gap: the field still lacks an exascale-trained framework that turns global historical EO priors into a practical capability for task-adaptive compression recovery with a ratio at the scale of several orders of magnitude.

\subsection{Compression Methods for Earth Observation}
%%drm: 体现一下超低码率下的遥感物理特性的丢失（不只是PSNR下降或者模糊，更重要的是传感器特性，如光谱、结构等）
%%zjx:DONE

Compression is essential for reducing the storage and transmission cost of large-scale Earth observation data. Existing methods, ranging from standardized codecs such as CCSDS~\cite{book2005image} and JPEG2000~\cite{gonzalez2009jpeg2000} to video-based adaptations built on HEVC~\cite{radosavljevic2020lossy} and VVC~\cite{seltsam2023adaptive}, have steadily improved coding efficiency for operational sensing systems. More recently, learned compression has further advanced rate-distortion performance by modeling image statistics and redundancy more effectively~\cite{balle2018variational,minnen2018joint}. Yet these methods still largely share the same basic formulation: they encode the current image and then attempt to reconstruct that same image mainly from the information retained in the transmitted bitstream.

At moderate bitrates, this formulation is often sufficient. In the ultra-low-bit-rate regime, however, the preserved signal becomes too limited to support faithful recovery from the bitstream alone. For Earth observation data, the resulting degradation is not only reflected in lower reconstruction quality or blurred textures, but more importantly in the loss of sensor-related physical characteristics, including spectral signatures, fine spatial structures, and cross-band relationships that are critical for downstream scientific analysis. In this regime, the central bottleneck is no longer only coding efficiency, but whether reconstruction can be supported by priors that extend well beyond the current observation.

This limitation has motivated the emergence of generative compression. In natural images, HiFiC~\cite{mentzer2020high} and later diffusion-based methods such as CDC~\cite{yang2023lossy} show that, at very low bitrates, high-quality recovery increasingly depends on combining compact representations with strong learned priors. Several recent remote sensing studies have begun to move in this direction. Earth+~\cite{du2025earth+} uses historical observations in a reference-assisted transmission framework, while MAGC~\cite{ye2025map} and COSMIC~\cite{zhang2024cosmic} explore generative reconstruction for remote sensing imagery. However, these methods still mainly recover the target from the current image plus limited nearby reference information, or focus primarily on RGB imagery. The current state of the art therefore still lacks a framework that treats global historical EO archives themselves as the primary prior source for ultra-low-bit-rate multispectral recovery.

%1.压缩方法->生成式方法先进性
%2. 遥感领域基础模型规模（全球数据+Trainable参数）
%3. 大规模训练性能（打表）

%第一类：传统压缩与遥感专用压缩
% 写它们在压缩率、是否支持全局先验、是否支持按需重建上的边界。
%
% 第二类：学习式压缩/生成式重建
% 写它们在 ultra-low bitrate、全局训练规模、工作流性能方面的边界。
% 第三类：大规模 AI/HPC 训练 SoA
% 这里可以引入近几年 Frontier、Alps、Aurora 上的 LLM/AI for science 相关 Gordon Bell finalist/获奖工作：这些工作展示了 exascale AI training 的可能性，但尚未把“全球科学数据生成式压缩”定义为一个独立 workload，也没有解决其特有的数据尺度、压缩 operating range 和 I/O/内存瓶颈。

\subsection{Global-Scale Remote Sensing Foundation Models}

Recent advances in remote sensing foundation models suggest that large-scale historical EO archives can indeed support the learning of strong priors. Existing efforts already cover a broad spectrum, including spectral modeling, global time-series pretraining, multi-sensor representation learning, multimodal spatio-temporal modeling, and language-conditioned EO generation, as represented by SpectralGPT \cite{hong2024spectralgpt}, Prithvi-EO-2.0 \cite{szwarcman2025prithvi}, DOFA \cite{xiong2024neural}, SkySense V2 \cite{zhang2025skysense}, OlmoEarth \cite{herzog2025olmoearth}, Text2Earth \cite{liu2025text2earth}, and RingMoE \cite{bi2025ringmoe}. Meanwhile, recent generative models such as MetaEarth \cite{yu2024metaearth} and TerraMind \cite{jakubik2025terramind} further suggest that these large-scale priors can be extended beyond representation learning toward global-scale image generation and multimodal generative modeling.  

Functionally, existing works can still be broadly grouped into two categories. The first consists of representation-oriented models, which mainly support understanding tasks such as classification, segmentation, and change detection by learning transferable EO representations from large archives. The second consists of generative models, which explore image generation, text- or multimodal-conditioned generation, and related EO completion tasks. Compact instruction-following vision-language extensions such as TinyRS-R1 \cite{koksal2025tinyrs} further broaden this landscape toward more general EO reasoning, but they are oriented more toward recognition and reasoning than toward recovery or reconstruction.

What these works establish is feasibility, not yet the specific capability required in this paper. They show that global historical EO archives can support strong prior learning, but they do not yet turn those priors into a framework for recovering a specific target observation under an extreme bitrate constraint from the small amount of information retained after compression. In other words, they demonstrate that global EO priors are learnable, but not yet that they are usable for compression-conditioned recovery.

Moreover, many recent EO foundation models span platforms or modalities through platform-aware branches, expert modules, or other modality-specific adaptation paths, as seen in DOFA, SkySense V2, OlmoEarth, and RingMoE. This is a sensible response to EO heterogeneity, but when the objective shifts to global-scale, multi-platform, multispectral compression recovery, it also implies substantial costs in training, storage, updating, and deployment. The state of the art therefore suggests that global prior modeling is possible, while still leaving open how to organize those priors into a practical, scalable framework for ultra-low-bitrate recovery.

\vspace{-0.4cm}

\subsection{Large-Scale Training Practice and Optimization}

\begin{table*}[h]
\centering
\caption{State of the Art Training Practices on Massively Parallel Supercomputer. MFU for Model FLOPs Utilization.}
\label{tab:soa}
\small % 减小字体到 small 或 footnotesize
\setlength{\tabcolsep}{8.2pt} % 稍微收紧列间距
\begin{tabular*}{\textwidth}{lccccccccccr}
\toprule
Application  &  Field  & Model Size & Batch Size & Hardware & Scale & MFU & Sustained PFLOP/s \\
\midrule
Narayanan et al. \cite{narayanan2021efficient} & Language  & 1000 B & 6.3 M & NVIDIA A100 & 3,072 GPUs & 52\% & 502 \\
MT-NLG \cite{smith2022using}  & Language  & 530 B & 4.0 M & NVIDIA A100 & 3,360 GPUs & 36\% & 379.7 \\
MegaScale \cite{jiang2024megascale} & Language  & 175 B & 12.5 M & NVIDIA A100 & 12,288 GPUs & 55\% & 2166.3 \\
Dash et al. \cite{dash2024optimizing}  & Language  & 1000 B & 19.7 M & AMD MI250X & 2,048 GPUs & 31.9\% & 188.0 \\
AxoNN \cite{singh2024democratizing}  & Language  & 60 B & 16.8 M & NVIDIA H100 & 6,144 GPUs & 23\% & 1423.1 \\
\midrule
MProt-DPO \cite{dharuman2024mprot}  & Protein Design & 3.5 B & 4.9 M & Intel Max 1550 & 38,400 GPUs & 45.5\% & 4110\textsuperscript{*} \\
ORBIT-2 \cite{wang2025orbit}  & Climate  & 10 B & 671 M & AMD MI250X & 65,536 GPUs & $\sim$16\% & 4100 \\
\rowcolor{gray!15}Tabuchi et al. \cite{tabuchi202116} & Cosmology & 7 M & 4 K & Fujitsu A64FX & 16384 CPUs & $\sim$2.0\% & $\sim$2.22 \\
\midrule
SkySense V2 \cite{zhang2025skysense} & Remote Sensing & 0.67 B & 4 M & NVIDIA H20 & 128 GPUs & / & / & \\
Prithvi-EO-2.0 \cite{szwarcman2025prithvi} & Remote Sensing & 0.6 B & 4 M & NVIDIA A100 & 240 GPUs & / & / & \\
OlmoEarth \cite{herzog2025olmoearth} & Remote Sensing & 0.3 B & 150 K & NVIDIA B200 & / & / & / & \\
RingMoE \cite{bi2025ringmoe} & Remote Sensing & 14.7 B & / & Ascend 910 & 512 NPUs & / & / & \\
\rowcolor{gray!15}\textbf{D2AR (This work)} & Remote Sensing & 6.3 B & 252 M & Armv9 LX2 CPU & 40,960 CPUs & 15.7\% & \textbf{1543} \\
\bottomrule
\end{tabular*}
\begin{flushleft}
\small \textsuperscript{*} RL training (DPO); reference model inference only, with no gradients or backpropagation. The two rows highlighted in grey represent works that rely on large-scale training on CPU platforms.
\end{flushleft}
\vspace{-0.3cm}
\end{table*}

Recent large-scale AI training studies, summarized in Table~\ref{tab:soa}, show that high sustained performance at scale depends on more than simply increasing parallelism. Across modern GPU and TPU systems, efficient training increasingly relies on the joint design of parallelization strategy, memory management, communication scheduling, and runtime execution. Frameworks such as Megatron-LM \cite{shoeybi2019megatron}, Megatron-DeepSpeed \cite{rasley2020deepspeed}, and AxoNN \cite{singh2022axonn} illustrate how hybrid parallelism and memory-aware training can improve throughput and scalability, while systems such as MegaScale \cite{jiang2024megascale} further highlight the importance of runtime robustness and system-level coordination in long-running jobs at extreme scale. Table~\ref{tab:soa} also shows that, although remote sensing foundation models are beginning to reach substantial scale, their reported training system characteristics remain far less explored than those of frontier language and scientific AI workloads.

However, the workload studied in this paper is not simply another instance of frontier model training. It combines global historical archive ingestion, large generative modeling, long visual sequences, multispectral physical constraints, and ultra-low-bitrate recovery objectives in a single system problem. Existing state-of-the-art results are therefore highly informative as performance references, but they do not directly resolve the challenge of making this workload practical. That challenge becomes even sharper on an emerging Armv9 CPU supercomputer, where many accelerator-oriented training recipes cannot be applied directly. In our setting, efficient execution is further constrained by limited SME utilization, higher runtime overhead, hierarchical HBM/DDR memory, and the interaction of multiple system factors that prevent the training stack from reaching high performance. The remaining gap is thus not only whether exascale training is possible, but whether this particular Earth observation workload can be trained efficiently enough to make global historical priors operational.

\section{Target Platform and Architectural Challenges: The Armv9-Based LineShine Supercomputer}

\subsection{System Architecture}

The LineShine supercomputer at the National Supercomputing Center in Shenzhen (NSCC-SZ) is an exascale HPC-AI converged system built on Armv9-based LX2 processors and interconnected by the LingQi high-speed network (LQLink) with 1.6~Tb/s per node.

\begin{figure}[h]
    \centering
    \includegraphics[width=0.48\textwidth]{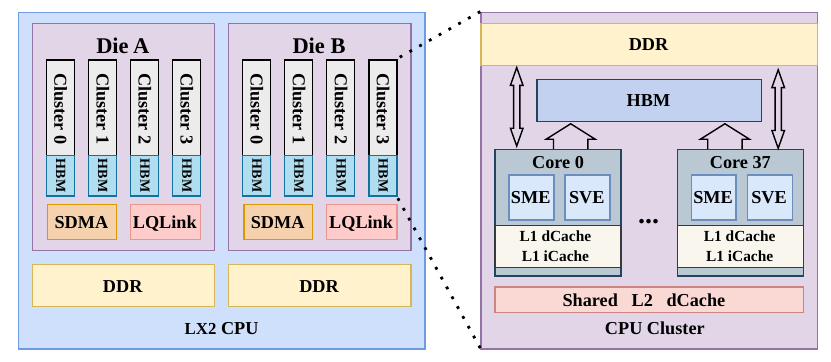}
    % \vspace{-0.6cm}
    \caption{Architectural overview of the LX2 processor.}
    \label{fig:lx2_arch}
\end{figure}

As shown in Figure~\ref{fig:lx2_arch}, each compute node contains two LX2 processors. Each processor integrates two compute dies and 304 CPU cores, organized into eight CPU clusters with 38 cores per cluster. Each core has a 32~KB L1 instruction cache and a 32~KB L1 data cache, while each cluster shares a 28.5~MB L2 cache.

LX2 adopts a hierarchical memory system that combines high bandwidth with large capacity. Each processor provides 32~GB on-package HBM with 4~TB/s peak aggregate bandwidth, together with 256~GB off-package DDR. Each die contains four cluster-local HBM domains and four DDR domains, giving 16 NUMA domains per processor. In this design, each cluster is associated with one HBM domain and one DDR domain. HBM access is strongly locality-sensitive, while DDR access is more uniform within a die but shared across multiple clusters. Each die also includes a dedicated SDMA engine for DDR-HBM data movement.

For computation, LX2 supports both Armv9 SVE and SME. SVE enables vector execution, while SME provides matrix acceleration through ZA tile registers and outer-product instructions, making it particularly effective for dense kernels such as GEMM. A single LX2 processor delivers a theoretical peak of 60.3~TFLOP/s in Float64, 240~TFLOP/s in BFloat16/Float16, and 960~TOP/s in INT8.

% \vspace{-0.3cm}

% \vspace{-0.2cm}
\subsection{Challenges and Opportunities}

Although LX2 offers strong compute capability and memory bandwidth, achieving efficient large-scale training on the LineShine supercomputer remains challenging.

The first challenge is the limited local memory available for training states. In large-scale training, parameters, activations, gradients, and optimizer states cannot be kept entirely within the memory resources local to a single cluster. On LX2 CPU, each cluster is associated with 4~GB HBM and 32~GB DDR, which is far from sufficient for the full working set of modern large-scale training. They must therefore be distributed across clusters, with DDR serving as the main storage space. This makes the organization of training states a basic requirement for execution and an important factor in later data access and communication cost.

The second challenge is the lack of an efficient software stack for large-scale CPU training. Existing deep learning frameworks and runtime systems are largely optimized for GPU platforms, where high-throughput kernels help amortize launch and scheduling overheads. On CPU platforms, these overheads become much more visible. Operator dispatch, runtime scheduling, tensor management, and synchronization can all take a non-negligible fraction of execution time. Efficient communication support for large-scale CPU training is also limited. Existing communication libraries are not well tuned for this setting, especially for BFloat16 communication, topology-aware collectives, and fine-grained overlap between communication and computation. As a result, communication overhead remains a major barrier to high sustained performance on LineShine.

Another challenge is topology-sensitive memory access. HBM delivers high bandwidth only when accessed locally. In our measurements, local HBM bandwidth reaches 450~GB/s, but drops to 230~GB/s for same-die remote access, 170~GB/s across dies, and 45~GB/s across CPUs. DDR bandwidth is more stable within a die, but still decreases from 125~GB/s locally to 110~GB/s across dies and 45~GB/s across CPUs. This behavior makes locality-aware data placement essential rather than optional.

The final challenge is kernel efficiency. While SME offers high peak throughput for matrix operators, sustaining that performance requires coordinated control of data layout, cache residency, memory movement, and kernel granularity. Training introduces frequent tensor reshaping, irregular memory access, and short-lived kernel execution, all of which can leave the matrix units underutilized if the runtime is not carefully co-designed.

Taken together, these challenges point to four key optimization opportunities on LineShine: parallel design for distributed training states, framework and communication support tailored for large-scale CPU training, locality-aware use of hierarchical HBM and DDR memory, and SME-oriented kernel and runtime co-design for high sustained efficiency.

\vspace{-0.4cm}
\section{Innovations Realized}
% \begin{figure*}
%     \centering
%     \includegraphics[width=\linewidth]{img/overall-scratch.pdf}
%     \caption{Caption}
%     \label{fig:placeholder}
% \end{figure*}
\begin{figure*}
    \centering
    \includegraphics[width=\linewidth]{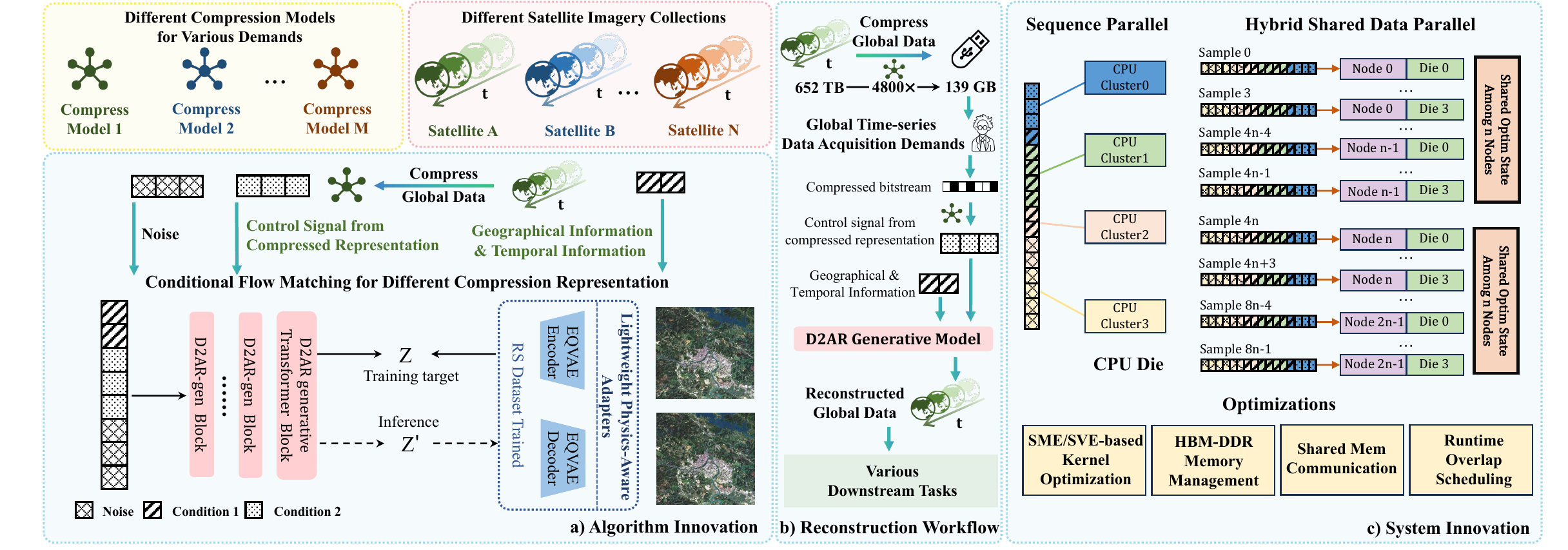}
    \vspace{-0.5cm}
    \caption{ Overview of the proposed historical-prior generative compression framework for Earth observation.
(a) Algorithm design for compressing global historical archives into generative priors with geographic and temporal conditioning.
(b) Reconstruction workflow for on-demand recovery from compressed representations for downstream use.
(c) System design for exascale training on the LineShine Armv9 CPU supercomputer.}
    \label{fig:overall-inno}
\vspace{-0.2cm}
\end{figure*}

\subsection{Overview}

This work introduces an integrated solution for global-scale Earth observation data compression and reconstruction from three tightly connected perspectives, as shown in Figure~\ref{fig:overall-inno}. At the framework level, we reorganize how data are compressed, delivered, and reconstructed for use. At the model level, we turn massive historical archives into controllable reconstruction capability. At the system level, we make this training paradigm practical on an exascale Armv9 CPU platform. 
% Taken together, these innovations reposition compression from a passive utility for storage and transmission into an active, task-adaptive mechanism for Earth observation data use.

\subsection{Framework-Level Innovations}
% (0.5pages)
%灵活压缩比、地理信息注入、物理一致性
%这里需要一个非常high level的结构图和

The primary motivation for our redesign stems from the extreme imbalance between the scale of Earth observation archives and the practicality of using them. To process global-scale data while maintaining physical fidelity, conventional monolithic pipelines face severe computational bottlenecks. At the same time, the heterogeneity of multi-source satellite platforms leads to non-standardized execution flows and massive I/O overhead. We propose a Dual-Decoupled Asymmetric Compression and Reconstruction (D2AR) framework that addresses this challenge at the system level by fundamentally reorganizing the coupling among compression, reconstruction, and user access. By doing so, it turns a fragmented scientific workflow into a structured computation-and-service pipeline that can simultaneously support extreme compression, flexible reconstruction, and downstream scientific use.

%The primary motivation for our algorithmic redesign stems from the extreme imbalance between data scale and computational resources in exascale Earth observation. To process global-scale archives and maintain physical fidelity, traditional end-to-end models confront a severe computational bottleneck where the heterogeneity of various satellite platforms leads to non-standardized execution flows and massive I/O overhead. On an exascale supercomputer, the system efficiency is highly sensitive to the consistency of computational kernels and the continuity of memory access. Therefore, we transform the traditional compression task into a standardized, two-stage generative workflow. By \hl{decoupling} \hl{sensor-specific} physical properties from the core generative backbone, we convert a \hl{fragmented scientific problem into a homogeneous computational workload that can fully exploit the massive throughput and hardware-software co-design of heterogeneous supercomputing nodes.}

\subsubsection{Task Decoupling and Asymmetric Space-Ground Computation}

%%%drm: add reference
To eliminate the inefficiency of training independent models for specific compression ratios and modalities, D2AR introduces a systemic task decoupling. The workflow is divided into two highly specialized execution stages. In the on orbit stage, the lightweight frontend D2AR-comp flexibly utilizes arbitrary existing compression models to transform diverse remote sensing inputs into a highly compressed intermediate bitstream. This bitstream is then processed by a lightweight pretrained feature mapper, which remaps it into control tokens representing the compressed signal. In the ground based stage, rather than treating these tokens as a direct pixel source for traditional decoding, they serve as a conditional control signal to guide D2AR-rec, our core Flow Matching reconstruction model. This profound architectural shift converts a data intensive transmission and decoding bottleneck into a highly homogeneous compute intensive generative task. For D2AR-rec, we adopt a generative model architecture based on Z-Image. Compared to existing generative models such as Stable Diffusion 3 and FLUX, which rely on massive dual stream architectures and incur severe memory bandwidth overheads, the Z-Image architecture utilizes a unified single stream diffusion transformer that significantly reduces parameter redundancy and memory access costs. This allows the ultra low entropy compressed signal to be progressively reconstructed with high fidelity within a model that encapsulates the global historical data priors of the specific sensor platform. Therefore, the supercomputer can concentrate its peak capability on a unified reconstruction logic while seamlessly supporting flexible on demand operations across extreme compression ratios.

\subsubsection{User-Centric Reconstruction Workflow}

This framework-level decoupling also changes how scientists interact with Earth observation data in practice. Instead of forcing users to acquire, move, and locally manage massive raw archives, supercomputing centers can expose compressed observation information at multiple precision levels as a new data service layer. Researchers then select the compression level most appropriate for their scientific target and invoke the corresponding D2AR-rec model for reconstruction and analysis. 
Empirically, Table 3 demonstrates that this framework achieves substantial improvements in high fidelity reconstruction over existing compression paradigms across extreme compression scenarios at magnitudes ranging from thousands to tens of thousands of times. Furthermore, Table 5 proves that thousandfold compression incurs negligible accuracy degradation in downstream land cover scene classification tasks, while ten thousandfold compression results in only a marginal performance drop of approximately 1\% in both F1 score and mean average precision.
In this workflow, storage, transfer, and reconstruction are no longer fixed upstream decisions; they become configurable elements of data use itself. This is the practical mechanism through which D2AR turns extreme compression into a new paradigm for accessing and using Earth observation data.

\subsection{Model-Level Innovations}

At the model level, D2AR separates sensor-specific physical heterogeneity from the core generative learning problem, while injecting global historical knowledge directly into the reconstruction path. This combination allows a unified reconstruction backbone to operate across multiple observation settings, effectively preserving physical consistency while maintaining a robust capability to recover information degraded under extreme compression.
%without sacrificing physical consistency or the ability to recover information lost under extreme compression.

\subsubsection{Physical Decoupling and Standardized Generative Backbone}
%%%drm: add reference, EQ-VAE
Recognizing the intrinsic physical heterogeneity of Earth observation sensors, D2AR implements a modular physical decoupling strategy. We use pre-trained EQ-VAE~\cite{eqvae_2025} encoders and decoders as lightweight physics-aware adapters. The unique imaging physics and spectral characteristics of specific sensors, such as synthetic aperture radar or multispectral platforms, are encapsulated within these adapters. Crucially, the optimization objective of D2AR-rec is not to reconstruct pixel-level images directly, but to generate the structured latent space produced by the EQ-VAE encoder. By executing Flow Matching~\cite{flowmatching_2022} within this latent domain, we decouple complex physical restoration from the core generative process. Once latent reconstruction is complete, the corresponding EQ-VAE decoder maps the result back to the original physical space. This design keeps the primary computational workload, namely the core backbone of D2AR-rec, standardized and platform-agnostic. By isolating platform-dependent physics in modular adapters, we establish a consistent execution flow across modalities and enable unified optimization of memory access patterns and communication kernels on exascale supercomputing nodes.

\subsubsection{Prior-Driven Reconstruction and Global Information Recovery}

To push beyond the theoretical reconstruction boundaries imposed by extreme compression ratios, D2AR-rec introduces a global geographic prior injection mechanism that transforms massive historical observation archives into actionable reconstruction capability. We encode global spatiotemporal priors into continuous geographic embeddings and dynamically inject them into the latent Flow Matching process as a global control. This establishes a deep semantic linkage between the current compressed observation and the global information accumulated by the corresponding sensor platforms. By leveraging this shared global context, the model can effectively compensate for the structural and spectral high frequency information that is lost during extreme compression. Empirically, Table 4 validates the effectiveness of the performance improvements achieved by scaling up the global data learning size. This joint global and local control strategy substantially unleashes the utility of massive historical datasets, effectively turning the storage capacity and memory bandwidth of the supercomputer into a direct gain in the physical fidelity of reconstructed observations.

\subsection{System Innovations}

Training large generative models on an emerging Armv9 CPU with hierarchical memory requires more than a conventional GPU-oriented stack. End-to-end efficiency is jointly constrained by three coupled factors: data reuse in SME-based kernels, limited HBM capacity across operators and tensor lifetimes, and the communication–computation balance shaped by system topology.

Our design follows a coordinated, layered approach from kernels to memory to parallelism. First, we develop an SME-oriented GEMM strategy that adapts tiling and buffering to problem geometry and cache limits. Second, we treat HBM as a global placement problem, prioritizing critical tensors under tight capacity constraints. Third, we co-design sequence and hybrid data parallelism with hardware topology to align communication and state placement with the system structure.

% Together, these innovations turn the system from a collection of local optimizations into a coordinated execution framework for efficient large-scale training on hierarchical-memory CPUs.
\subsubsection{Reuse-Directed Asymmetric SME-GEMM}

On Armv9 CPUs with hierarchical memory, GEMM performance is not only limited by compute power, but also by how well data stays in cache and gets reused. If cache locality is poor, even compute-heavy workloads quickly become memory-bound. To address this, we do not use a single fixed kernel. Instead, we adapt the computation strategy based on the shape of the problem, choosing the approach that maximizes data reuse. 

Specifically, we develop two complementary SME micro-kernels, \(64\times K\) by \(K\times16\) and \(16\times K\) by \(K\times64\), and select between them at runtime according to the relative sizes of \(M\) and \(N\). This design enables two dual dataflows: one organized around M-major reuse and the other around N-major reuse. In each case, the wide-tile operand is kept thread-private to maximize intra-thread reuse, while the narrow-tile operand is streamed across the computation.

To ensure data stays in cache, we derive the parameter \texttt{magic\_k} based on how much data can fit in L2 cache, rather than tuning it empirically. We further use reuse-aligned thread partitioning and near-uniform tiling to reduce tail imbalance and synchronization overhead under non-divisible problem sizes. Concretely, the partitioned dimension follows the selected reuse direction, so that threads operate on tiles with stable private reuse and minimized interference.

\begin{figure}
    \centering
    \includegraphics[width=\linewidth]{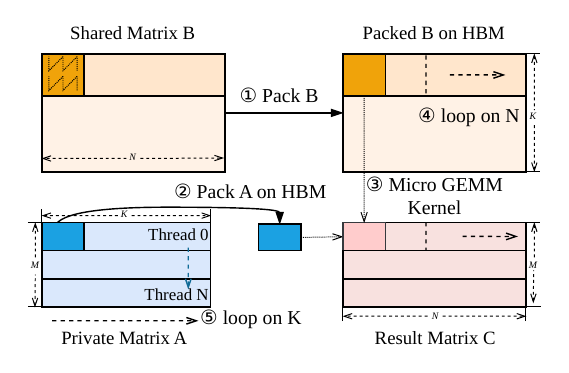}
    \vspace{-0.7cm}
    \caption{GEMM workflow. Threads cooperatively pack the shared matrix B into HBM. The computation is organized as a tiled loop over the K dimension: each thread packs its local tile of A into HBM, traverses the shared dimension, and performs micro-kernel computations to accumulate results into matrix C.}
    \label{fig:gemm-opt}
    \vspace{-0.6cm}
\end{figure}

We further apply asymmetric memory optimizations, as shown in Figure \ref{fig:gemm-opt}. The streamed operand is prefetched and stored in a large HBM buffer to sustain high bandwidth, while the thread-private operand uses only small local buffers without prefetching. For the output matrix, we adopt a K-aware accumulation policy: when \(K<\texttt{magic\_k}\), the result is written back directly; otherwise, FP32 partial sums are accumulated in HBM before final write-back. Bias addition and output scaling are fused into the main compute and write-back path to further reduce memory traffic.

Overall, this design treats GEMM optimization as a unified problem of data reuse and locality. Kernel shape, thread mapping, memory layout, and accumulation strategy are all co-designed to maximize cache reuse and minimize memory movement.

\subsubsection{Operator- and Lifetime-Aware HBM Strategy}

While the SME-GEMM optimization improves the efficiency of core dense computation, overall training performance remains strongly constrained by the limited capacity of on-package HBM. On the target platform, HBM cannot hold all model states and intermediate tensors simultaneously, so its usage must be planned globally rather than assigned uniformly. We therefore formulate HBM placement as a constrained resource-allocation problem: under a fixed capacity budget, HBM should be reserved for tensors whose placement yields the highest end-to-end performance benefit after accounting for both kernel sensitivity and tensor lifetime.

\begin{figure}[h]
    \centering
    \includegraphics[width=0.48\textwidth]{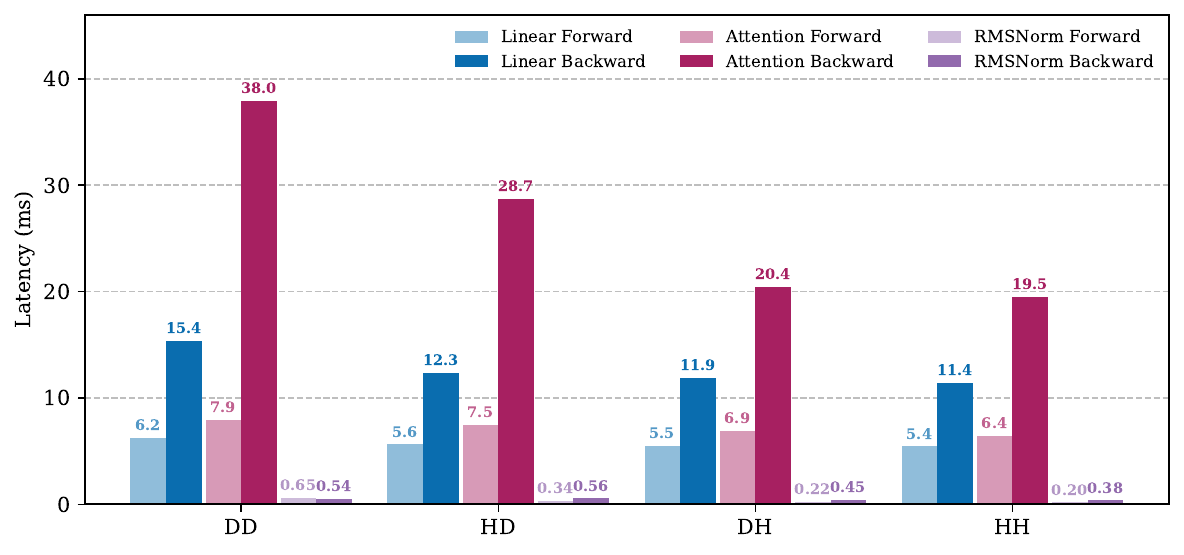}
    \caption{Forward and backward latency of core operators across different memory configurations (ms). \texttt{H} denotes HBM and \texttt{D} denotes DDR. The two-letter column headers indicate the memory placement of input and output tensors, respectively. For instance, \texttt{HD} denotes that input tensor resides in HBM while output tensor is allocated in DDR.}
    \label{fig:hbm-ddr-op}
    % \vspace{-0.6cm}
\end{figure}

Our strategy is guided by the operator-level characterization in Figure~\ref{fig:hbm-ddr-op}. The results show that the benefit of HBM placement is highly non-uniform across operators and execution phases. In the forward pass, the attention operator exhibits the clearest sensitivity to HBM bandwidth, while other operators obtain smaller gains in the context of full-model execution. Based on this observation, we place only the output activations of attention in HBM during the forward pass, while keeping other intermediate tensors in DDR to relieve HBM pressure. This choice concentrates scarce HBM capacity on the most bandwidth-sensitive forward tensors instead of diluting it across all operators. Moreover, because the attention outputs are subsequently cast from float32 to float16, the corresponding HBM allocation can be released immediately after the cast, further reduce the HBM footprint.

In the backward pass, the situation differs. Backward kernels consistently benefit from HBM, and most gradients and intermediates are short-lived, so we place all backward allocations in HBM. The main exception is parameter gradients, which persist for accumulation and communication. To avoid exhausting HBM, we offload them to DDR during the SP communication phase, freeing space while retaining the benefits of the HBM.

In conclusion, our strategy considers both \emph{operator sensitivity}, which captures the performance gain from using HBM, and \emph{tensor lifetime}, which determines whether this gain is sustainable under limited capacity. This joint design prioritizes HBM for the most critical operators, maximizing overall training throughput.

\subsubsection{Asynchronous Runtime and Overlap Execution}

To reduce runtime overhead and improve hardware utilization, we design an asynchronous execution runtime. We introduce a dedicated launch thread to overlap framework overheads (e.g., operator dispatch, memory allocation, and tensor view operations) with kernel execution. Our profiling shows that the default PyTorch runtime incurs approximately 24.2\% overhead. With the asynchronous runtime, this overhead is reduced to 1.9\%, resulting in near-optimal overlap between computation and framework execution. In addition, we implement communication-computation overlap using hook-based scheduling and explicit synchronization mechanisms. All communication operations are overlapped with ongoing computation whenever possible, minimizing idle time and improving overall throughput. Furthermore, we assign dedicated CPU cores to handle communication tasks, reducing interference with computation.

\subsubsection{Topology-Aligned Parallelization Strategy}

% After optimizing kernel execution and HBM usage, the remaining challenge is to scale training across the full machine without introducing excessive communication or memory overhead. 
To scale efficiently to full machine, we must carefully design the parallelization strategy that matches the hardware topology, rather than directly adopting a generic distributed design. We therefore use a hierarchical parallelization scheme aligned with the processor structure. Each MPI process is bound to one CPU cluster, which contains 38 cores, with 37 cores available to the application, along with 32\,GB DDR and 4\,GB HBM. Within a die, four processes in different NUMA domains can communicate through shared memory.

On top of this shared-memory, we apply sequence parallelism (SP) across the four intra-die processes and use ring attention for attention computation. This reduces per-process activation memory and keeps SP communication within the low-latency shared-memory. At a larger scale, we introduce hybrid shared data parallelism (HSDP), which partially shares and shards optimizer states across processes within four nodes. Compared with full replication, this reduces optimizer-state memory and eliminates redundant optimizer computation.

We further optimize communication inside the SP domain. Shared memory is used to accelerate both ring data exchange in attention and gradient all-reduce within the SP group. In addition, we adopt different communication schedules for forward and backward. The forward pass exchanges \(K\) and \(V\), while the backward pass exchanges \(Q\) and \(\mathrm{grad\_out}\). Compared with a symmetric design that exchanges \(K\)- and \(V\)-related tensors throughout backpropagation, this reduces backward communication volume by about 50\%.

% Overall, the design follows a simple principle: each form of sharing and communication is placed on the hardware domain where it is most efficient. Parameter sharing is confined within a die, SP communication is restricted to the shared-memory, and optimizer-state sharding is extended across nodes only when it provides net memory and computation benefits. This makes SP and HSDP work as a unified hierarchical parallel framework rather than two independently stacked techniques.

\begin{figure}[h]
    \centering
    \includegraphics[width=0.48\textwidth]{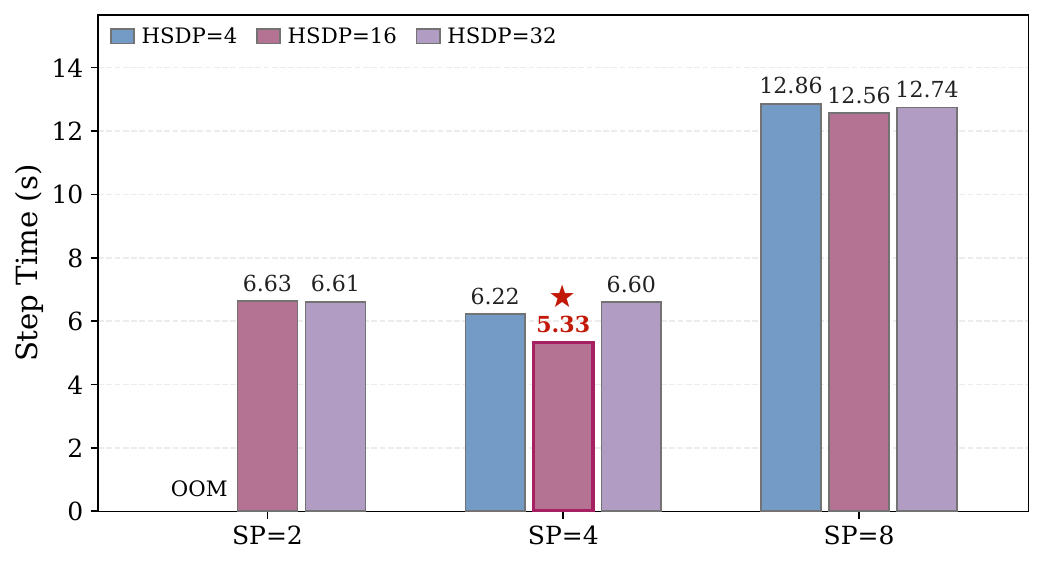}
    \caption{Performance comparison across different parallelism configurations for D2AR-rec-6B on 128 nodes, including Hybrid Shared Data Parallelism (HSDP) and Sequence Parallelism (SP).}
    \label{fig:bestconf}
\end{figure}

This design integrates SP and HSDP into a unified hierarchical parallel framework. We further evaluate the impact of SP and HSDP group sizes, as shown in Figure~\ref{fig:bestconf}. Specifically, we test two sets of configurations: SP$=4$ with HSDP$=4/16/32$, and HSDP$=16$ with SP$=2/4/8$. Here, SP$\times$HSDP denotes the total number of shards for each configuration. The results are used to understand the trade-off between SP communication overhead and HSDP memory savings, and to select the final parallel configuration.

\section{How Performance was Measured}
% \subsection{Model Architecture Details}

% \begin{table}[h]
% \caption{ARCHITECTURAL CONFIGURATIONS OF THE Z\_Image MODELS USED IN THE PERFORMANCE EXPERIMENTS.}
% \label{tab:operator_time}
% \centering
% \setlength{\tabcolsep}{5pt}
% \begin{tabular}{lcccc}
% \toprule
% \textbf{Model} & \textbf{Parameters} & \textbf{Layers} & \textbf{Hidden Size} & \textbf{Heads} \\
% \midrule
% Z\_Image-3B & 3B & 17 & 3840 & 30 \\
% Z\_Image-6B & 6B & 36 & 3840 & 30 \\
% Z\_Image-8B & 8B & 52 & 4096 & 32 \\
% \bottomrule
% \end{tabular}
% \end{table}

We use three D2AR-rec model configurations in the performance experiments, namely D2AR-rec-3B, D2AR-rec-6B, and D2AR-rec-8B, as summarized in Table~\ref{tab:operator_time}.
\begin{table}[h]
\vspace{-0.1cm}
\caption{ARCHITECTURAL DETAILS IN THE PERFORMANCE EXPERIMENTS.}
\label{tab:operator_time}
\centering
\small
\setlength{\tabcolsep}{5pt}
\begin{tabular}{lcccc}
\toprule
\textbf{Model} & \textbf{Parameters} & \textbf{Layers} & \textbf{Hidden-Size} & \textbf{Heads} \\
\midrule
D2AR-rec-3B    & 3B & 17 & 3840 & 30 \\
D2AR-rec-6B    & 6B & 36 & 3840 & 30 \\
D2AR-rec-8B    & 8B & 46 & 4096 & 32 \\
\bottomrule
\end{tabular}
\vspace{-0.3cm}
\end{table}

% \subsection{Measurements}

We report two complementary performance metrics. The primary metric is the sustained end-to-end BFloat16 FLOP/s over full training iterations, capturing the realized cost of computation, communication, and data movement in actual distributed execution. Alongside this, we report Model FLOPs Utilization (MFU), which quantifies the fraction of the system's theoretical peak FLOPs that is effectively achieved during training. In addition, we provide a compute-only metric, referred to as Peak FLOPs in the following figures and tables, which is derived solely from forward and backward model computation and excludes communication and I/O overhead. 
% Together, these metrics disentangle intrinsic algorithmic compute capability from overall end-to-end system efficiency.

For each configuration, we run the training loop for twenty iterations with fixed model size, sequence length, and batch configuration, and report the average per-iteration time over the last ten iterations to reduce warmup effects and capture steady-state execution. We follow the analytical FLOP formulation used in Megatron-LM \cite{shoeybi2019megatron} to estimate model computation, and compute sustained BFloat16 FLOP/s by dividing the analytical model FLOPs by the measured end-to-end iteration time. MFU is then defined as the ratio between the measured sustained BFloat16 FLOP/s of the full system and the aggregate theoretical BFloat16 peak of the participating nodes. Since a single LX2 node provides 480~TFLOP/s theoretical BFloat16 peak, all MFU values in this work are normalized against this vendor-advertised theoretical peak rather than an empirically measured GEMM upper bound.

% We adopt this normalization because, on this emerging architecture, neither open-source nor vendor-provided BLAS libraries yet provide a credible empirical upper bound across the relevant workload shapes. Therefore, unlike prior studies on mature GPU platforms \cite{singh2024democratizing, shoeybi2019megatron}, we do not normalize by an empirically measured GEMM peak. The resulting MFU should thus be interpreted as a conservative efficiency metric relative to hardware peak capability. To provide additional context, we include the vendor BLAS implementation in the single-node comparisons as an implementation baseline, but not as the machine upper bound.

\section{Performance Results}
% include scalability (weak and strong), time to solution, efficiency (of bottleneck resources), and peak performance (2 pp max)743564

We evaluate performance results from three complementary perspectives. First, we quantify how much each optimization contributes on a single node. Second, we test whether these gains are preserved at scale and whether the proposed system can sustain exascale training throughput on the full machine. Third, we validate that the training system supports the intended scientific objective of historical-prior generative compression across extreme compression ratios, global prior utilization, and downstream task utility.

\subsection{Single Node}
\begin{figure}[t]
    \centering
    \includegraphics[width=0.48\textwidth]{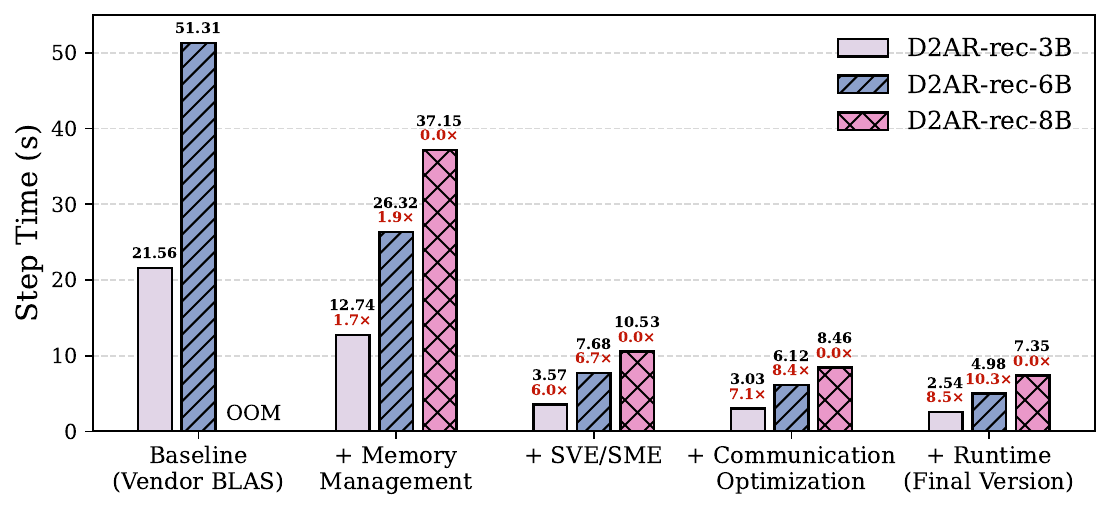}
    \caption{Single-node runtime under different optimization stages across model sizes.}
    \label{fig:stepwise}
\end{figure}

% \subsection{Single Node}
We first study single-node performance to isolate the effect of local optimizations before moving to distributed scaling. Figure~\ref{fig:stepwise} reports the runtime under successive optimization stages for three model sizes, D2AR-rec-3B, D2AR-rec-6B, and D2AR-rec-8B. 

% Each group in the figure corresponds to one optimization stage, and the three bars within a group show the runtimes of the three models.
% We first evaluate the proposed system on a single LX2 node to evaluate the impact of local optimization before moving to distributed scaling. Fig.~\ref{fig:stepwise} reports the runtime under different optimization stages for three model sizes, D2AR-rec-3B, D2AR-rec-6B, and D2AR-rec-8B. In the figure, each group corresponds to one optimization stage, and the three bars within each group show the runtime of the three models.

% As shown in Fig.~\ref{fig:stepwise}, the proposed optimizations consistently reduce runtime for all three models. Starting from the baseline implementation based on vendor BLAS, introducing the HBM-aware memory strategy already yields a clear reduction in step time, indicating that memory placement is an important factor on CPU platform with hierarchical memory. After further incorporating the optimized SVE/SME kernels, runtime drops more substantially, showing that improving dense computation efficiency becomes critical once the main memory bottleneck is alleviated. Communication optimization and the final asynchronous runtime then bring additional gains by reducing exposed communication and synchronization overhead.
As shown in Figure~\ref{fig:stepwise}, the proposed optimizations consistently reduce runtime for all three models. Starting from the vendor-BLAS baseline, introducing the HBM-aware memory strategy already yields a clear reduction in step time, indicating that memory placement is a first-order factor on this hierarchical-memory CPU platform. After further incorporating the optimized SVE/SME kernels, runtime drops more substantially, showing that dense-compute efficiency becomes the dominant lever once the main memory bottleneck is alleviated. Communication optimization and the asynchronous runtime then bring additional gains by reducing exposed communication, launch, and synchronization overhead.

% For the 6B model, the runtime is reduced from 51.31\,s in the baseline to 26.32\,s after HBM-aware memory management, then to 7.68\,s with SVE/SME kernel optimization, 6.12\,s with communication optimization, and finally 4.98\,s with the asynchronous runtime, corresponding to an overall speedup of 10.3$\times$. Similar stepwise improvements are observed for the 3B models, with final version reduced from 21.56\,s to 2.54\,s. For the 8B model, the baseline configuration results in an out-of-memory error, while execution becomes feasible after applying memory management optimizations.
For the 6B model, the runtime is reduced from 51.31\,s in the baseline to 26.32\,s after HBM-aware memory management, then to 7.68\,s with SVE/SME kernel optimization, 6.12\,s with communication optimization, and finally 4.98\,s with the asynchronous runtime, corresponding to an overall speedup of 10.3$\times$. Similar stepwise improvements are observed for the 3B model, whose runtime decreases from 21.56\,s to 2.54\,s. For the 8B model, the baseline configuration results in an out-of-memory failure, while execution becomes feasible once the memory optimizations are introduced.

% Two observations can be made from these results. First, the optimization trend is consistent across model sizes, which suggests that the proposed design is not tied to a specific model configuration. Second, the relative contribution of different optimization stages is also stable: HBM-aware placement removes a substantial part of the bandwidth bottleneck, SME kernel optimization delivers the largest compute-side gain, and communication/runtime optimization further improves end-to-end efficiency by reducing non-compute overhead. Overall, these results show that single-node performance on the target platform depends on coordinated optimization of computation, memory, communication, and runtime execution, rather than on any single component alone.
% Two observations follow from these results. First, the optimization trend is consistent across model sizes, which suggests that the proposed design is not tied to a single model configuration. Second, the relative contribution of the stages is also stable: HBM-aware placement removes a substantial part of the bandwidth bottleneck, SME kernel optimization delivers the largest compute-side gain, and communication/runtime optimization further improves end-to-end efficiency by reducing non-compute overhead. Overall, the single-node results show that high performance on the target platform emerges from coordinated optimization of computation, memory, communication, and runtime execution rather than from any single technique in isolation.

\subsection{Weak Scaling Performance}
\subsubsection{Weak Scaling Efficiency}
\begin{figure}[t]
    \centering
    \includegraphics[width=0.48\textwidth]{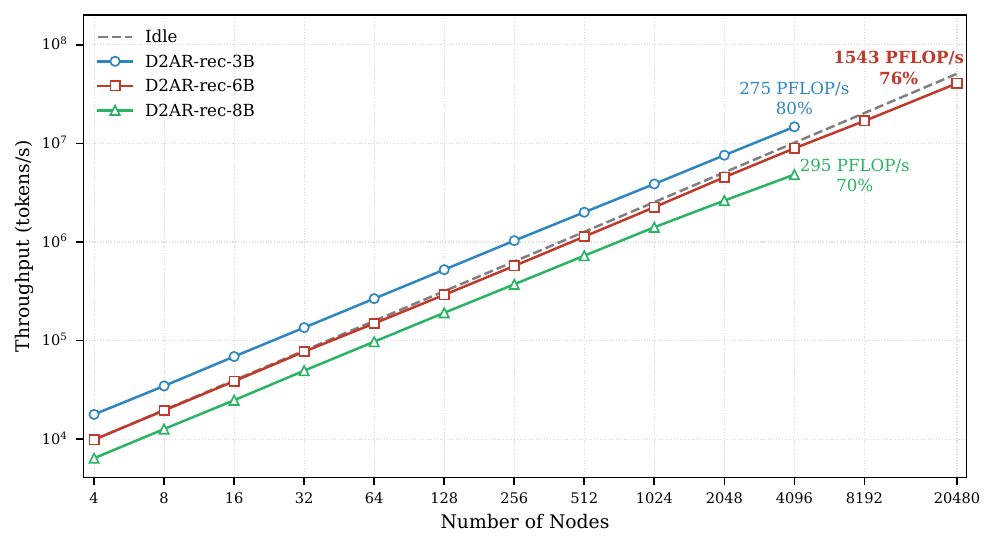}
    \caption{Weak scaling performance of the proposed system for three model sizes. In all cases, the global workload is increased proportionally with the number of nodes while keeping the per-node workload fixed.}
    \label{fig:weak-scaling}
\end{figure}

We evaluate weak scaling by proportionally increasing the global workload with the number of nodes while keeping the per-node workload fixed. Figure~\ref{fig:weak-scaling} reports the results for three model sizes: 3B model from 1 to 4096 nodes, 6B model from 1 to 20480 nodes, and 8B model from 1 to 4096 nodes. 
% This setting measures whether the proposed training framework can preserve efficiency as the system expands to larger machine sizes.

Overall, all three configurations exhibit good weak scaling behavior. The 6B model delivers the main large-scale result, sustaining 1.54~EFLOP/s at 20480 nodes with a weak scaling efficiency of 76.0\%, demonstrating that the proposed system remains effective even at full-machine scale. Under this configuration, the corresponding Peak FLOPs reaches 2.16~EFLOP/s. For the 3B and 8B models, the system scales to 4096 nodes and achieves 275~PFLOP/s and 295~PFLOP/s sustained performance, with corresponding weak scaling efficiencies of 80.4\% and 70.4\%, respectively. 

% These results indicate that the proposed design is not tied to a single model size, but generalizes across a meaningful range of model configurations.

% The overall trend of the three curves also reveals the interaction between scale and model size. The 3B model reflects the scaling behavior of a relatively lighter configuration, where communication and runtime overheads become visible earlier as the system expands. The 8B model, in contrast, carries higher per-step computation and therefore benefits more from computation amortization, although its larger parameter and activation footprint also places greater pressure on memory management and communication. The 6B model provides a balanced operating point, showing that the proposed system can maintain strong scalability while handling a practically important model size at extreme machine scale.

% These weak scaling results confirm that the combination of topology-aware parallelization and asynchronous communication-computation overlap effectively controls the overheads associated with large-scale execution. As node count increases, the system continues to translate additional hardware resources into sustained training throughput rather than being dominated by communication, data movement, or runtime imbalance. This behavior is critical for enabling global-scale training workloads on the target supercomputer.

% \subsubsection{Full-Scale Training Scenario}

\subsubsection{Full-Scale Training Scenario}

Unlike strong scaling, where more nodes are used to reduce the iteration time of a fixed small problem, our goal is to continuously incorporate larger portions of the global historical archive while keeping the per-node workload, optimizer behavior, and memory footprint stable. In this regime, additional nodes increase total training throughput and global batch size, so that larger machine scale translates into broader archive coverage rather than simply faster processing of a fixed dataset fragment.

% This distinction is particularly important for historical-prior generative compression. A single epoch over global Earth observation archives is already a massive computational task, and the scientific value of the model depends on exposing it to increasingly broad geographic and temporal coverage. Therefore, the key full-machine question is not whether a fixed task can be further accelerated, but whether training remains efficient as the global workload grows with machine size. The weak-scaling results in Figure~\ref{fig:weak-scaling} show that this is indeed the case, with the proposed system maintaining 76\% efficiency at 20480 nodes. More importantly, full-scale training is beneficial not only from the systems perspective but also from the optimization perspective. Starting from the same pretrained checkpoint and continuing training on global Earth observation data, the full-machine large-batch setting reaches the same intermediate loss level within only 50 steps, whereas the 4-node configuration still does not reach that level after more than 5000 steps. The large-batch trajectory is also visibly smoother, with smaller loss oscillation throughout training. These results show that the role of the exascale machine in this work is not simply to accelerate a conventional training recipe, but to enable a qualitatively different regime in which the model can ingest more of the Earth, more efficiently and more stably, to build the desired historical prior.

This distinction is especially important for historical-prior generative compression. A single epoch over global Earth observation archives is already a massive computational task, and the model’s scientific value depends on exposure to broader geographic and temporal coverage. Accordingly, the key full-machine question is not whether a fixed task can be further accelerated, but whether training remains efficient as the global workload grows with machine size. Figure~\ref{fig:weak-scaling} shows that this is indeed the case, with the proposed system maintaining 76\% weak-scaling efficiency at 20480 nodes.
More importantly, full-scale training improves not only system throughput but also optimization. Starting from the same pretrained checkpoint and continuing training on global Earth observation data, the full-machine large-batch setting reaches the same intermediate loss level in only 50 steps, whereas the 4-node configuration still does not reach that level after more than 5000 steps. The large-batch trajectory is also smoother, with smaller loss oscillations throughout training. Together, these results show that the exascale machine is not used merely to accelerate a conventional training recipe, but to enable a different training regime in which the model can ingest more of the Earth, more efficiently and more stably, to build the desired historical prior.

\subsection{Application-Level Validation}

\subsubsection{Generalization for Different Compression Ratios}
To evaluate the flexibility and robustness of this paradigm, we evaluate three cases using distinct compression baselines: HL-CompNet~\cite{hl_decom_2024} (Case 1), LIC-TCM~\cite{liu2023learned} (Case 2), and MLIC++~\cite{jiang2025mlicpp} (Case 3), spanning compression ratios from 6000$\times$ to 18,000$\times$. Using Sentinel-2 multispectral imagery as a benchmark, we conduct ablation studies across eight consecutive years of data from the global top-1,000 cities. Table~\ref{tab:method-comparison} shows that the generative reconstruction model consistently outperforms signal-only recovery baselines across all tested compression regimes. The gains are especially clear on perceptual and structural metrics such as LPIPS and MS-SSIM, while also improving or preserving physically relevant indicators such as NDVI error. 
These results validate that the decoupled framework remains effective across diverse and extreme operating points rather than only at a single bitrate.

% To evaluate the flexibility and robustness of this paradigm, we integrate three distinct compression baselines (HL-CompNet~\cite{hl_decom_2024}, LIC-TCM~\cite{liu2023learned}, and MLIC++~\cite{jiang2025mlicpp}) with compression ratios ranging from 6000$\times$ to 18,000$\times$. Using Sentinel-2 multispectral imagery as an example, we conduct extensive ablation studies across eight consecutive years of data from the global top-1,000 cities. The results demonstrate that our D2AR-rec approach consistently outperforms existing signal-only recovery baselines across all tested compression regimes. Notably, the model achieves substantial gains in sensor-specific physical fidelity. Spectral indicators such as the Mean Absolute Error of Normalized Difference Vegetation Index (NDVI MAE) and structural metrics including Multi-Scale Structural Similarity (MS-SSIM) and Learned Perceptual Image Patch Similarity (LPIPS) show significant and consistent improvements. These results validate the efficacy of our decoupled framework in handling highly diverse and extreme-scale Earth observation scenarios.

\begin{table}[h]
\caption{Reconstruction Performance Across Different Compression Ratios.}
\label{tab:method-comparison}
\renewcommand{\arraystretch}{1.2}
\resizebox{\columnwidth}{!}{
\begin{tabular}{lccccc}
\toprule
% \textbf{Method} & \textbf{Comp. Ratio} & \textbf{BPP$\downarrow$} & \textbf{PSNR$\uparrow$} & \textbf{LPIPS$\downarrow$} & \textbf{MS-SSIM$\uparrow$} & \textbf{NDVI$\downarrow$} \\
\textbf{Method} & \textbf{Comp. Ratio} & \textbf{PSNR$\uparrow$} & \textbf{LPIPS$\downarrow$} & \textbf{MS-SSIM$\uparrow$} & \textbf{NDVI$\downarrow$} \\
\midrule
\addlinespace[2pt]

Case 1~\cite{hl_decom_2024}   &  6104  & 13.2252  & 0.6497 & 0.5922 & 0.0982 \\
+D2AR-rec (Ours)              &  6104  & 14.9484  & 0.1074 & 0.9238 & 0.0915 \\
Case 2~\cite{liu2023learned}  &  6295  & 15.3820  & 0.4504 & 0.9087 & 0.0776 \\ 
+D2AR-rec (Ours)              &  6295  & 16.5998  & 0.1218 & 0.9537 & 0.0818 \\
Case 3~\cite{jiang2025mlicpp} &  17777 & 13.1023  & 0.6189 & 0.8416 & 0.0978 \\
+D2AR-rec (Ours)              &  17777 & 14.1299  & 0.1193 & 0.9018 & 0.0893 \\
\bottomrule
\end{tabular}
}
\end{table}
% psnr 13.9732
% LPIPS 0.2235
% MS-SSIM 0.9061
% NDVI 0.0862

\subsubsection{The Necessity of Utilizing Global Data}

To verify the necessity of comprehensive global data for effective prior learning, we conduct a comparative study focused on the density of spatial priors. We partition the dataset from 1,000 representative global cities into distinct configurations to observe the impact of geographic coverage on reconstruction fidelity. The control group uses the full extent of the 1,000 cities over a six-year period for training, while the experimental group intentionally restricts the spatial coverage to 900 cities over the same duration. Both models are evaluated on a held-out validation set comprising the remaining 100 cities during the subsequent two-year period. As shown in Table~\ref{tab:data-scale-comparison}, the model trained on the full geographic extent consistently outperforms the spatially restricted counterpart. This gap confirms that the generative recovery process is sensitive to the completeness of the global prior: adding the remaining 100 cities provides geographic context that improves generalization to unseen temporal intervals in those regions. These findings support a central claim of the paper, namely that using the full scale of global historical archives is not a mere data expansion, but a direct source of reconstruction capability.

% To verify the necessity of comprehensive global data for effective prior learning, we conduct a rigorous comparative study focused on the density of spatial priors. We partition our dataset from 1,000 representative global cities into distinct configurations to observe the impact of geographic coverage on reconstruction fidelity. The control group utilizes the full extent of the 1,000 cities over a six-year period for training, while the experimental group intentionally restricts the spatial coverage to 900 cities over the same duration. Both models are evaluated on a held-out validation set comprising the remaining 100 cities during the subsequent two-year period. The experimental results demonstrate that the model trained on the full geographic extent consistently outperforms the spatially-restricted counterpart. This disparity confirms that the generative recovery process is highly sensitive to the completeness of the global prior. Specifically, the inclusion of the additional 100 cities in the training phase provides critical geographic context that enhances the model ability to generalize to unseen temporal intervals in those regions. These findings provide empirical evidence that utilizing the full scale of global historical archives is a fundamental necessity rather than a mere data expansion for achieving high-fidelity reconstruction under extreme-scale Earth observation constraints.

\begin{table}[h]
\centering
\small
\caption{Ablation Study Validating the Performance Gains from Expanding Global Prior Coverage.}
\label{tab:data-scale-comparison}
\begin{tabular}{lcccc}
\toprule
\textbf{Data Scale} & \textbf{PSNR$\uparrow$} & \textbf{LPIPS$\downarrow$} & \textbf{MS-SSIM$\uparrow$} & \textbf{NDVI$\downarrow$} \\
\midrule
900 Cities & 12.1516 & 0.3273 & 0.8474 & 0.0934 \\
1000 Cities & 12.4708 & 0.2990 & 0.8586 & 0.0906 \\
\bottomrule
\end{tabular}
\end{table}

% \subsubsection{Scaling Behavior and Large Batch Optimization on Supercomputers}

% To effectively utilize the parallel computing capacity of exascale supercomputers and process large-scale global datasets, we conduct scaling experiments focusing on large-batch optimization. Empirical training loss trajectories show that expanding the global batch size mitigates fluctuations in the loss landscape. Specifically, the optimization trajectory of the large-batch configuration exhibits lower variance compared to the small-batch baseline. However, as training progresses, the marginal utility of scaling the batch size gradually diminishes, manifesting as a stabilized loss reduction rate rather than a continuous sharp decline. Training a generative foundation model on full-scale Earth observation data for a single epoch presents a substantial computational challenge. By leveraging the scalable architecture of heterogeneous supercomputing clusters, our large-batch strategy maintains training stability and convergence efficiency. Consequently, this approach makes the comprehensive processing of global-scale historical archives practically feasible without compromising optimization robustness.

\subsubsection{Downstream Utility}

To verify the practical value of extreme compression for downstream analysis, we conduct validation on a representative land-cover scene classification task using the DynamicEarthNet~\cite{toker2022dynamicearthnet} Sentinel-2 dataset, which contains 896 images with corresponding 7-class land-cover labels. Using the same dataset split, we establish strictly paired evaluation pipelines. For each compression baseline including HL-CompNet, LIC-TCM, and MLIC++, we train a ResNet 18~\cite{he2016deep} model exclusively on the reconstructed data and compare it against a corresponding ResNet 18 baseline independently trained on the original data.
%and a counterpart ResNet-18 trained and tested exclusively on the reconstructed data generated by our framework. 
We report Exact Accuracy~\cite{zhang2013review}, Macro F1-Score~\cite{sokolova2009systematic}, and mean Average Precision (mAP)~\cite{everingham2010pascal}. Exact Accuracy is a strict instance-level metric that counts a prediction as correct only when the entire multi-label vector matches the ground truth.

As shown in Table~\ref{tab:downstream-task}, the downstream classification performance on reconstructed data exhibits only limited degradation even under extreme compression. Across the tested settings, Macro F1 remains unchanged or changes only marginally, while mAP remains close to the original-data baseline. These results indicate that the proposed reconstruction pipeline preserves a substantial fraction of the semantic and structural information required by downstream tasks, rather than optimizing only for pixel-level fidelity.

\begin{table}[h]
\caption{Downstream task evaluation on the land cover scene classification.}
\label{tab:downstream-task}
\centering
\renewcommand{\arraystretch}{1.2}
\setlength{\aboverulesep}{0pt}
\setlength{\belowrulesep}{0pt}
\resizebox{\columnwidth}{!}{
\begin{tabular}{l | cc | cc | cc}
\toprule
\textbf{Method} & \multicolumn{2}{c|}{\textbf{Case 1 + D2AR-rec}} & \multicolumn{2}{c|}{\textbf{Case 2 + D2AR-rec}} & \multicolumn{2}{c}{\textbf{Case 3 + D2AR-rec}}  \\
\midrule
Metric & Original & Ours  & Original & Ours  & Original & Ours  \\
\midrule
Exact Acc   & 0.5000 & 0.4792 
& 0.4688 & 0.5000 
& 0.5000 & 0.4792 
\\
Macro F1 & 0.7540 & 0.7540
& 0.7540 & 0.7540 
& 0.7540 & 0.7547 
 \\
mAP       & 0.8535 & 0.8465
& 0.8761 & 0.8372 
& 0.8535 & 0.8312 
 \\
\bottomrule
\end{tabular}
}
\end{table}

\subsubsection{Global Compression and Reconstruction Application}

\begin{figure*}
    \centering
    \includegraphics[width=\linewidth]{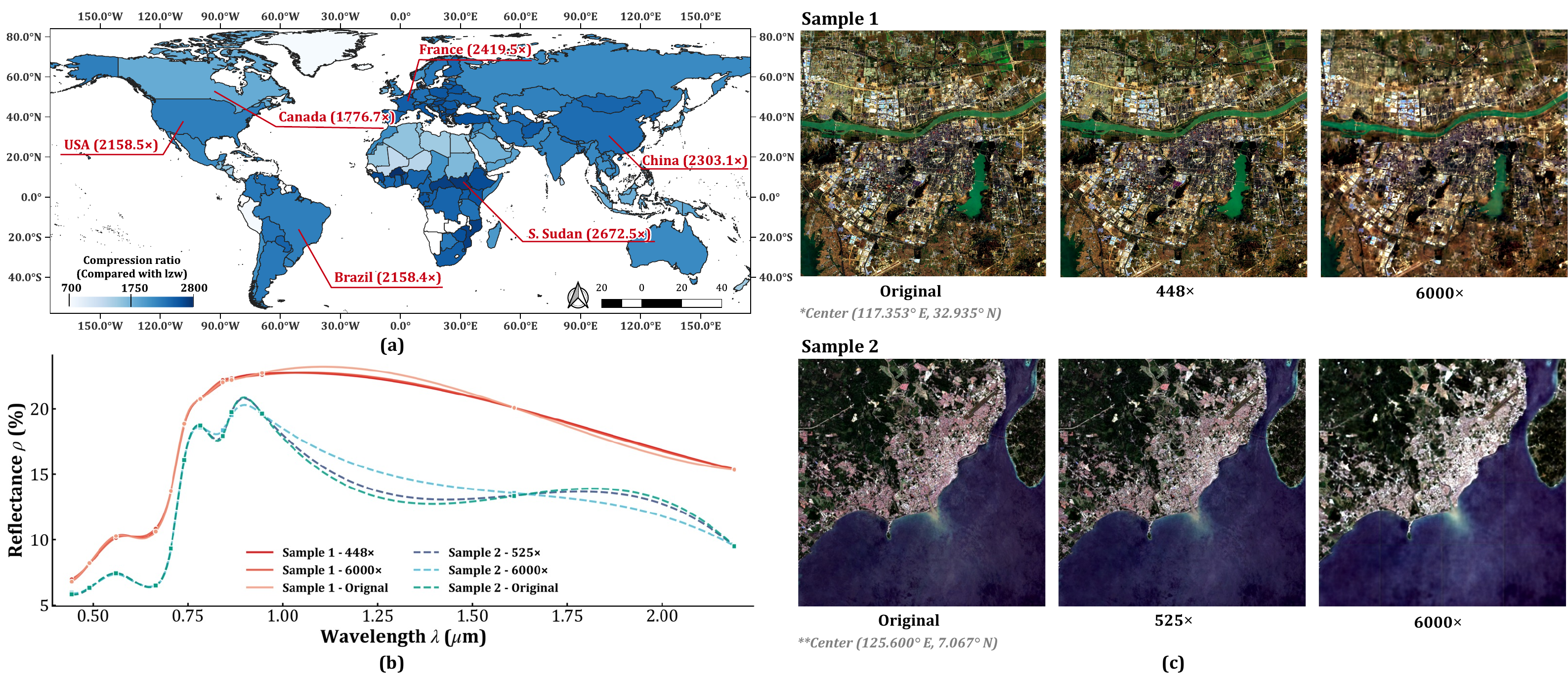}
    \vspace{-0.75cm}
    \caption{Visual Comparison under different compression ratios and corresponding spectral curves.}
    \label{fig:vis-recon}
\end{figure*}

Taking the global thousandfold compression of Sentinel 2 data as an example, Figure \ref{fig:vis-recon} (a) visualizes the relative gains of the lossy paradigm over traditional lossless methods. The spatial distribution reveals significantly higher compression gains in dense urban areas. Since high intrinsic information entropy renders lossless algorithms largely ineffective, the deep entropy reduction of the lossy frontend maximizes efficiency while severely exacerbating reconstruction challenges. This firmly validates the necessity of introducing a global prior driven generative model to synthesize discarded details.

% \begin{figure}
%     \centering
%     \includegraphics[width=\linewidth]{img/compress_res.pdf}
%     \caption{Global spatial distribution of compression ratios achieved by the proposed model on Earth observation data compared with lossless LZW method.}
%     \label{fig:global-comprs}
% \end{figure}

Figure~\ref{fig:vis-recon} (b) evaluates the visual reconstruction and Figure~\ref{fig:vis-recon} (c) shows the corresponding spectral curves across hundredfold to ten thousandfold compression scales. At hundredfold compression, D2AR delivers high fidelity reconstruction with highly consistent visual textures. Under extreme ten thousandfold compression, although micro textures are inevitably smoothed, the model successfully maintains the macroscopic semantic layout. Crucially, global physical priors effectively maintain the structural integrity of the spectral curves, ensuring high reliability for downstream quantitative analyses despite extreme information loss.

\section{Implications}

Continuous Earth observation is essential for addressing global challenges, yet sensing capabilities now vastly outpace downlink and storage capacities. Current pipelines treat each acquisition as an isolated file, ignoring the immense spatiotemporal redundancies of repeatedly measuring the same evolving planet. To resolve this bottleneck, our work suggests moving beyond merely building better codecs or larger archives by fundamentally redefining the role of historical data itself.
Specifically, we demonstrate that global historical archives can be condensed into actionable generative priors. Activated by ultra-low-bitrate features, these priors recover physical information absent from the transmitted signal. This transforms compression from a passive storage utility into an active mechanism for data utilization. In this paradigm, accumulated prior knowledge becomes as critical as transmitted bits, fundamentally allowing ground-based compute and historical knowledge to be systematically traded for satellite-to-ground bandwidth, storage footprint, and data accessibility.

This shift has implications across the full Earth observation ecosystem. On satellites, for platforms with sufficient onboard compute, the role of transmission can move from pixel-by-pixel relay toward task-adaptive feature export, quality-aware filtering, and progressive information release. At ground stations, reception can evolve from simple data ingress into the first scheduling layer of a space-ground computing continuum, where streams are routed according to latency, importance, and reconstruction objectives. At the level of the largest centralized archives, supercomputing data centers can transition from passive warehouses into generative prior engines that continuously absorb global history, update reconstruction models, and serve feature-based access at scale. At regional or institutional data centers, the same paradigm enables task-specific reconstruction, local adaptation, and selective caching without requiring full replication of petabyte-scale raw archives. For end users, especially those outside elite storage environments, access can gradually shift from downloading massive physical files toward obtaining compact features, derived products, or on-demand reconstructions matched to specific scientific tolerances. In this sense, the paradigm has the potential to restructure the entire data ecology of Earth science, from onboard acquisition to personal scientific use. Importantly, this is not a purely hypothetical direction: onboard AI missions have already begun filtering or compressing imagery before downlink, while major Earth observation platforms are increasingly combining archive access with cloud-side processing close to the data.

% Furthermore, a key architectural implication of our work is that the physical complexity of multi-source Earth observation data can be translated into an advantage for system scalability through a scientific decoupling strategy. We encapsulate the heterogeneous imaging physics of specific sensors within lightweight perception modules while standardizing the core generative inference within a unified backbone network. This not only ensures the physical consistency of the reconstructed data but also transforms fragmented platform adaptation tasks into highly homogeneous computational workloads on supercomputing nodes, achieving highly efficient computational throughput. This deep system and physics decoupling provides an excellent paradigm for interdisciplinary collaboration: Earth system scientists can focus on defining physical boundaries and exploring spatiotemporal patterns, while computer scientists can concentrate on overcoming system-level bottlenecks in storage bandwidth and computation scheduling.
The system implications are equally important. Condensing long-term global observation information into computational priors is not only a modeling problem but also an infrastructure problem dominated by data ingestion, preprocessing, scheduling, storage access, and distributed runtime efficiency. Our results therefore provide evidence that CPU-centered supercomputers, and Arm platforms in particular, can serve as highly effective substrates for large-scale scientific AI. Beyond raw dense computation, such systems offer natural advantages for the control-heavy and data-intensive parts of the workload, including complex preprocessing, distributed input/output orchestration, memory management, communication scheduling, and coupling with large storage systems. These properties are especially relevant for emerging hybrid scenarios in which model training, retrieval, simulation, tool use, and agent-like runtime logic coexist within the same AI-for-Science pipeline.

% Looking ahead, as global satellite constellations undergo explosive deployment, the profound gap between the exponential surge in orbital data generation and the slow growth of physical communication channels will persist. To resolve this dilemma, our research proposes and validates a Space-Ground Computing Continuum vision from a system perspective. The profound significance of this vision lies not only in breaking the ceiling of a single transmission link but also in restructuring the entire big data ecology of Earth science from the bottom up.
% Another important implication is methodological. The physical complexity of multi-source Earth observation does not necessarily have to reduce system efficiency. Through scientific decoupling, we isolate sensor-specific physics inside lightweight perception modules while standardizing the core generative computation inside a unified backbone. This not only preserves physical consistency across heterogeneous modalities, but also transforms fragmented scientific adaptation problems into a more homogeneous computational workload. More broadly, this provides a promising model for interdisciplinary collaboration: domain scientists can continue to define physical constraints, error tolerances, and scientific objectives, while computer scientists optimize the shared computational substrate on which these constraints are expressed and enforced.

Another implication is methodological. The physical complexity of multi-source Earth observation need not reduce efficiency. By decoupling, sensor-specific physics is handled in lightweight perception modules, while core generative computation is unified in a shared backbone. This preserves cross-modality physical consistency and turns fragmented adaptation into a more homogeneous workload. It also supports interdisciplinary collaboration: domain scientists define constraints and objectives, while computer scientists optimize the shared computational substrate on which these constraints are expressed and enforced.

% First, it is poised to profoundly evolve how satellites acquire and transmit data. Computationally constrained on-orbit nodes will no longer be limited to inefficient pixel-by-pixel physical transfers. Instead, they will distill redundant original observation fields into ultra-low-entropy intrinsic feature streams on demand, thereby significantly alleviating the communication pressure on satellite-to-ground channels.
Finally, the broader significance of this work may extend beyond Earth observation. We do not expect this paradigm to apply uniformly to all scientific data. However, it is likely to be most valuable in domains with four properties: repeated observation of the same or closely related physical system, very large data volume, strong structural redundancy across time or instruments, and downstream use patterns that rarely require unrestricted access to the entire raw archive at full fidelity. Earth observation is an especially strong instance of this regime, but similar opportunities may emerge in other scientific settings such as large sky surveys, climate and environmental monitoring systems, long-horizon simulation archives, and certain high-throughput imaging workflows. In fact, several large scientific infrastructures are already moving toward science-ready products, regional access centers, and next-to-data analysis rather than unrestricted raw-data distribution to every end user. From this perspective, historical-prior generative compression should be viewed not only as a new Earth observation technique, but as a candidate systems paradigm for future scientific data infrastructures.

In summary, this work points toward a new scientific computing pathway in which historical data are no longer treated merely as stored records, but as executable prior knowledge. By trading large-scale ground computation and long-term learned priors for bandwidth, storage, and data movement, the proposed space-ground collaborative infrastructure opens a practical route toward more efficient, more scalable, and more democratic use of Earth observation data. If sustained and generalized, this paradigm could help redefine not only how remote sensing data are compressed, but how major scientific data systems are acquired, distributed, and ultimately used.

% implications for future systems and applications (1 p max)

% \begin{acks}
% \end{acks}

\bibliographystyle{ACM-Reference-Format}
\bibliography{sample-base}

% \appendix


\end{document}